\documentclass{article}

\PassOptionsToPackage{numbers, compress}{natbib}

\usepackage[final]{neurips_2022_ml4ps} 

\usepackage[utf8]{inputenc} 
\usepackage[T1]{fontenc}    
\usepackage{hyperref}       
\usepackage{url}            
\usepackage{booktabs}       
\usepackage{amsfonts}       
\usepackage{nicefrac}       
\usepackage{microtype}      
\usepackage{xcolor}         
\usepackage{amsmath}
\usepackage{amssymb}
\usepackage{listings}
\usepackage{pgfplots}
\usepackage{tikz}
\usepackage{amsthm}
\usepackage{multirow}
\usepackage{mathtools}
\usetikzlibrary{external}
\usepackage[inline]{enumitem}

\usepackage{subcaption}

\newcommand\mass{$\boldsymbol{m}$}

\title{Decorrelation with Conditional Normalizing Flows}
\author{%
  Samuel Klein \\
  University of Geneva\\
  \texttt{samuel.klein@unige.ch} \\ 
  \And
  Tobias Golling \\
  University of Geneva\\
  \texttt{tobias.golling@unige.ch} \\ 
}

\begin{document}

\maketitle

\begin{abstract}
  The sensitivity of many physics analyses can be enhanced by constructing discriminants that preferentially select signal events. Such discriminants become much more useful if they are uncorrelated with a set of protected attributes.
  In this paper we show that a normalizing flow conditioned on the protected attributes can be used to find a decorrelated representation for any discriminant.
  As a normalizing flow is invertible the separation power of the resulting discriminant will be unchanged at any fixed value of the protected attributes.
  We demonstrate the efficacy of our approach by building supervised jet taggers that produce almost no sculpting in the mass distribution of the background.
\end{abstract} 

\section{Introduction}

To study particular signal processes many analyses in high energy physics (HEP) will generate both signal and background samples using a simulator and define a discriminant for preferentially selecting signal \cite{baldi2014searching, de2016jet, cms2017identification, butter2018deep, moreno2020interaction, ATL-PHYS-PUB-2017-017}. Once a discriminant has been constructed it is used in an event selection to create a signal enriched data sample.
In a HEP analysis this selection will only ever be one step in a chain of procedures.
A common downstream task is the study of the invariant mass spectrum, which is of interest because it typically falls smoothly for the background but not the signal.
The discriminant is invariably correlated with the invariant mass and this typically means the smooth background assumption is violated in the signal enriched sample.
Therefore the discriminant is required to satisfy smoothness criteria with respect to the changes it induces in the mass distribution of the background.

In this work we introduce a method that removes the mass correlation from a discriminant after it has been constructed and leaves its separation power unchanged at every mass value.
This method can be used to decorrelate any number of variables from any number of protected attributes and is complementary to other decorrelation methods. 
We study supervised machine learning models on a supervised task and demonstrate the proposed method significantly improves the decorrelation performance of existing approaches. 

\section{Method}

The method introduced here will be described as a procedure for decorrelating discriminants, but it can be applied to decorrelate any set of variables from any set of protected attributes, though this may come with a cost (App.~\ref{app:task_agnostic_decor}).
When combined with a machine learning model the methods we present can be seen as additional layers that calibrate the output of the model to enforce uniform background rejection as a function of the protected attributes.


\paragraph{Existing approaches}
Existing methods often perform decorrelation during the construction of the discriminant. 
This can be done by minimizing a measure of the correlation between the discriminant $s$ and the protected attributes \mass{} while training a machine learning model~\cite{disco, mode, Shimmin_2017, Louppe_adversarial}. 
The construction of such a discriminant is performed by minimizing the expectation over some objective function of the form
\begin{equation}
    \centering
    \mathcal{L} = \mathcal{L}_{\textrm{class}}(s_{\phi}(\boldsymbol{x}), y) + \alpha \mathcal{L}_{\textrm{decor}}(s_{\phi}(\boldsymbol{x}), \boldsymbol{m}),
    \label{eq:mode_like}
\end{equation}
where $\phi$ parameterizes the discriminant and $y$ is the class label for the sample described by features $\boldsymbol{x}$.
Minimizing the loss function $\mathcal{L}_{\textrm{class}}$ maximizes the utility of the discriminant while minimizing $\mathcal{L}_{\textrm{decor}}$ reduces the correlation between $\boldsymbol{m}$ and $s_{\phi}(\boldsymbol{x})$. 
The parameter $\alpha$ controls the relative contribution of each of the loss functions and is the minimal additional parameter that can appear in such loss functions, though more parameters can appear in $\mathcal{L}_{\textrm{decor}}$.

Decorrelation methods that use Eq.~\ref{eq:mode_like} require additional work when building the discriminant.
Both moment decomposition (MoDe)~\cite{mode} and distance correlation (DisCo)~\cite{disco} methods have some small overhead in terms of training, but only introduce the single hyper parameter $\alpha$ into the optimization. 
All of these methods can be composed with the decorrelation method we propose. Other approaches are detailed in App.~\ref{app:existing_approach}.

\paragraph{Decorrelation with conditional normalizing flows}
A normalizing flow~\cite{tabak_flows} is an invertible map between two distributions -- a data distribution from which we have samples and a base distribution for which we know the likelihood. A conditional normalizing flow $p_\theta$ can approximate a data distribution $p_D(s(\boldsymbol{x}) | \boldsymbol{m})$ by defining a neural network $f_\theta(s(\boldsymbol{x}), \boldsymbol{m})$ that is invertible given \mass{} and a base distribution $p$ that is independent of \mass. The model is fit to data by maximizing the log-likelihood under the change of variables formula, 
\begin{equation}
    \log p_\theta(s(\boldsymbol{x}) | \boldsymbol{m}) = \log p ( f_\theta(s(\boldsymbol{x}), \boldsymbol{m}) ) + \log \left| \det J_{f_\theta}(s(\boldsymbol{x}), \boldsymbol{m} 
    ) \right|,
    \label{eq:decor_eq}
\end{equation}
where $J_{f_\theta}$ is the Jacobian of $f_\theta$.
If this function is fit perfectly the distribution of $f_\theta(s(\boldsymbol{x}), \boldsymbol{m})$ will be the same as $p$ and therefore independent of the mass by definition. 
In this case $f_\theta(s(\boldsymbol{x}), \boldsymbol{m})$ will be a decorrelated representation of $s(\boldsymbol{x})$ with the same signal separation power as $s(\boldsymbol{x})$ at every value of \mass{} and therefore $f_\theta(s(\boldsymbol{x}), \boldsymbol{m})$ can be used in place of $s(\boldsymbol{x})$.

In the case of discriminants the function $f_\theta$ is required to be monotonically increasing to ensure the ordering remains unchanged at all values of \mass{}. 
In one dimension this condition can always be explicitly enforced post training of a flow using reflections of $f_\theta$ about the center of $p$ as discussed in App.~\ref{app:monotonic_flows}. 
As $f_\theta$ is an invertible transformation with positive slope, the discriminatory power of $s$ and $f_\theta$ will be the same for every value of $\boldsymbol{m}$.

\paragraph{Decorrelation with quantile regression}
Quantile regression~\cite{quantile_regression} is the task of finding the quantiles of a given distribution as a function of explanatory variables. 
Typically we know what quantiles we want to use for a given discriminant before it is constructed~\cite{CwolaOG}, and by learning these quantiles as a function of \mass{} we can use the conditional quantiles to perform event selections such that the distribution of $\boldsymbol{m}$ does not change.
Regressing quantiles is easier than training a conditional flow which has to learn the entire conditional cumulative distribution function.
The definition of quantiles was recently generalised to greater than one dimension~\cite{Carlier_2020,chernozhukov2017monge} and for decorrelating discriminants in more than one dimension this is the formulation that should be used to preserve the ordering that makes the discriminant useful.
This is discussed further in App.~\ref{app:quantile_regression}.

\section{Experiments}

\begin{figure}
    \centering
    \includegraphics[width=0.45\textwidth]{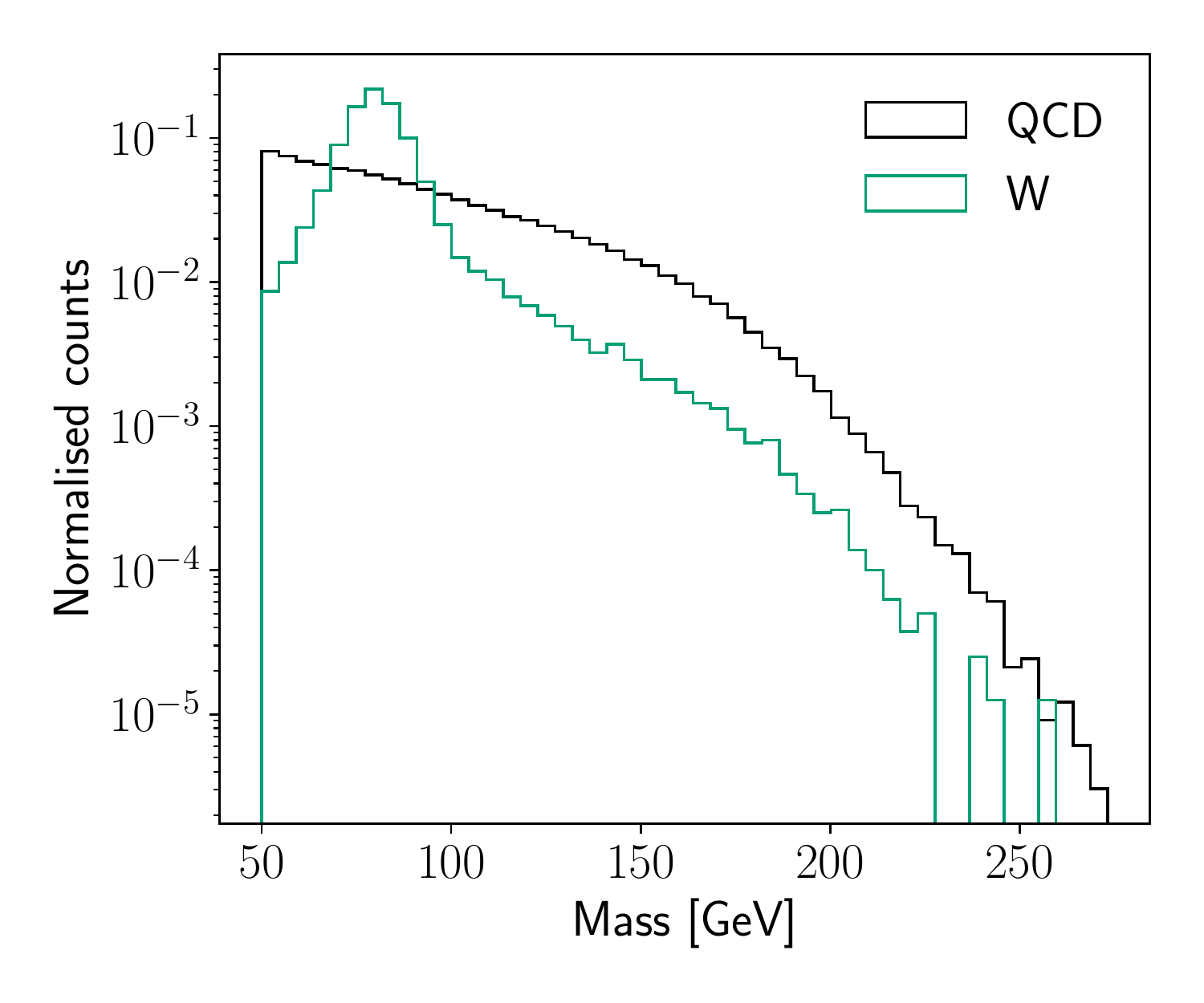}
    \caption{The invariant mass distribution for the background samples without applying any selection.}
    \label{fig:mass_only}
\end{figure}

In this section we will demonstrate the efficacy of the proposed method on a supervised classification problem.
The task is to separate multijet background events from hadronically decaying boosted \textit{W} bosons. 
The samples we use were provided by Refs~\cite{disco, gregor_kasieczka_2020_3606767} to emulate those used in an earlier ATLAS study on mass decorrelation techniques~\cite{ATL-PHYS-PUB-2018-014}. 
The same set of ten jet substructure variables as used in these studies were used as features to train the classifier. 
Both signal and background were generated by \textsc{Pythia} at $s=\sqrt{13}$ TeV with a detector simulated by \textsc{Delphes}. Jets are reconstructed using \textsc{FastJet}~\cite{cacciari2012fastjet,cacciari2006dispelling} and clustered using the anti-\textit{kt} algorithm~\cite{cacciari2008anti} with $R = 1.0$. Each jet is required to have transverse momentum $p_T \in [300, 400]$ GeV and mass  $m \in [50,300]$. For each jet ten substructure variables are calculated and used as input to the machine learning classifiers, these variables are the same as those used in previous studies of decorrelation~\cite{ATL-PHYS-PUB-2018-014, mode, disco}. The invariant mass distribution is shown in Fig.~\ref{fig:mass_only}. 
For this task the invariant mass is the protected attribute $\boldsymbol{m}$. 

\paragraph{Model settings}
For MoDe and DisCo models the discriminant is parameterized by a DNN with three hidden layers with $64$ nodes in each hidden layer and \textsc{Swish} activations~\cite{swish_activation} with a batch normalization layer after the first fully connected layer and a sigmoid activation on the output. This is the same model as used in the experiments of MoDe~\cite{mode}.
We also consider a more standard classifier with the same structure but with \textsc{ReLU} activations~\cite{relu} and no batch normalization, this will be referred to as a vDNN. 
The training procedure of the classifiers used in this paper is different to the trainings in the MoDe and DisCo papers and leads to classifiers that are more difficult to decorrelate.
In particular the classifiers we use have the learning rate annealed to zero during training and are trained without reweighting the $p_T$ distribution.
However, the decorrelation method developed here can be used with any training pipeline, in particular on top of the output of publicly available code such as that provided by DisCo~\cite{disco}, the result of this is shown in App.~\ref{app:disco_reproduced}.
All conditional normalizing flows are constructed with rational quadratic splines~\cite{durkan2019neural} and trained on the background only. Further training details are provided in App.~\ref{app:model_settings}.

\begin{figure}
    \centering
    \includegraphics[width=\textwidth]{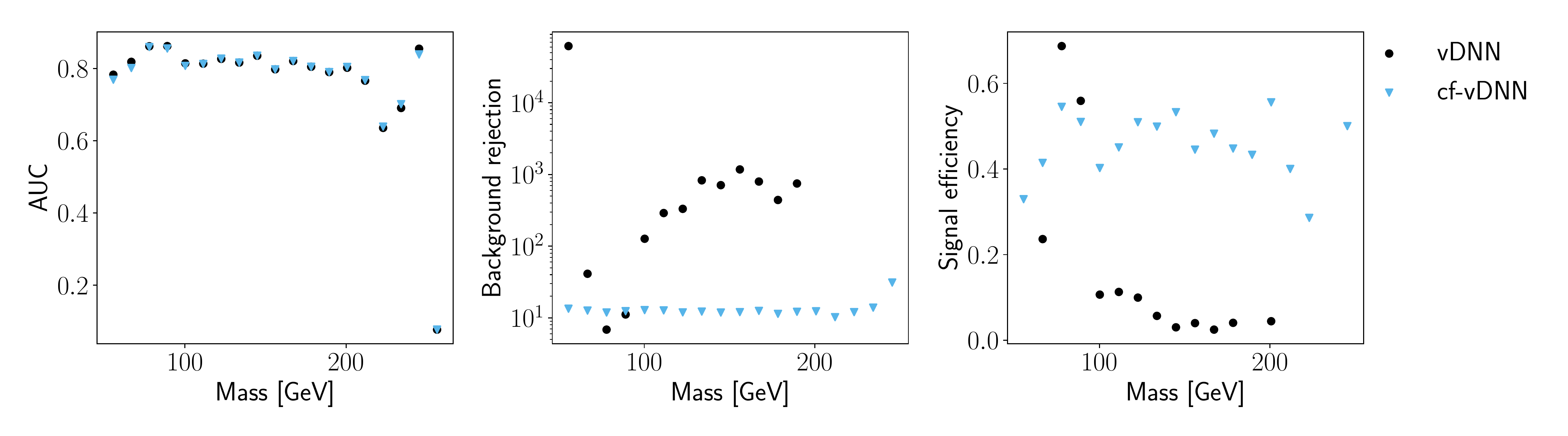}
    \caption{The AUC, background rejection and signal efficiency in bins of mass for a vDNN and the discriminant constructed by a conditional normalizing flow using the output of the vDNN (cf-vDNN).
    The binned rejection and efficiency are calculated using a threshold on the mass inclusive discriminant that gives $50\%$ signal efficiency.
    In the case of the vDNN some high mass bins do not have enough statistics to calculate background rejection and signal efficiency at an inclusive $50\%$ signal efficiency threshold.
    }
    \label{fig:auc_mass} 
\end{figure}

\paragraph{Signal separation}
A conditional flow decorrelated vDNN model (cf-vDNN) has the same separation power in bins of mass and modulates the background and signal mass distributions less than a vDNN with no decorrelation as shown in Fig.~\ref{fig:auc_mass}. 
The difference in background rejection for mass inclusive thresholds means that the signal separation using these thresholds is different. 
This is reflected in the mass inclusive AUC which is $0.90$ for a vDNN and $0.84$ for the cf-vDNN.
This is discussed further in App.~\ref{app:signal_separation}.

\begin{figure} 
    \centering
    \includegraphics[width=\textwidth]{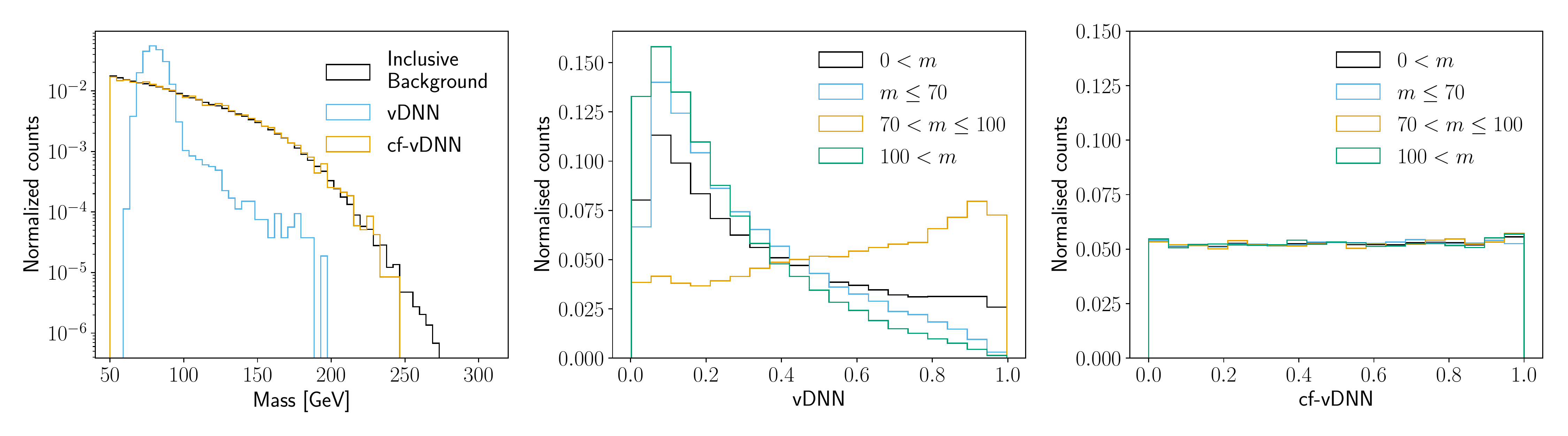}
    \caption{
        The invariant mass distribution for the background samples without applying any selection and a threshold rejecting $50\%$ of the signal for different discriminants applied to the background (left) and the distribution of the discriminant over QCD (background) samples for a vDNN (middle) and a conditional flow trained on the vDNN output (cf-vDNN, right) in bins of the mass.
    } 
    \label{fig:correlation_demo}
\end{figure}

\paragraph{Decorrelation}
The output distribution of a vDNN classifier trained without decorrelation changes significantly in different mass bins as shown in Fig.~\ref{fig:correlation_demo}.
In contrast the cf-vDNN model has approximately the same distribution in all mass bins. Using the vDNN output directly results in preferentially selecting samples in a certain mass range, while the cf-vDNN output exhibits no such preference.

To compare the conditional flow decorrelation to existing methods we study the inverse of the Jensen-Shannon divergence $1/\mathrm{JSD}_{50}$ and background rejection power (inverse false positive rate) $R_{50}$ at $50\%$ signal efficiency. The $\mathrm{JSD}$ is computed between samples that pass and fail the selection.
These metrics are correlated in the regime of finite statistics with an optimal trade off that can be estimated from data by applying random selections with the same proportions defined by $R_{50}$, we will refer to this optimal tradeoff as ideal. 

\begin{figure}
    \centering
    \includegraphics[width=\textwidth]{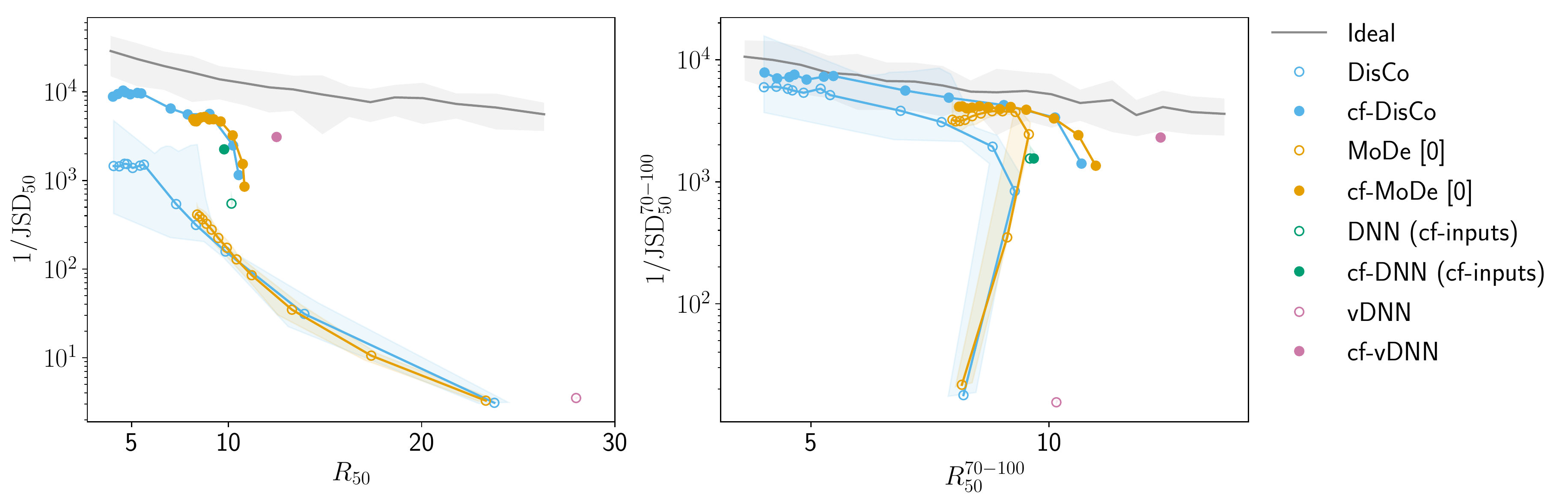}
    \caption{
        The inclusive background rejection at $50\%$ signal efficiency ($R_{50}$) and the inverse of the Jensen-Shannon divergence between the mass distribution of events that do and do not pass the cut at the same threshold ($1 / \mathrm{JSD}_{50}$) in the left plot.
        The background rejection in the mass window $[70-100]$ GeV at $50\%$ inclusive signal efficiency ($R_{50}^{70-100}$) and the inverse Jensen-Shannon divergence in the same mass bin ($1 / \mathrm{JSD}_{50}^{70-100}$).
        The filled circles denote a discriminant constructed using a conditional normalizing flow.
        The shaded bands capture the maximum and minimum across ten different trainings. 
        Models with the cf prefix have been decorrelated with a conditional normalizing flow.
    } 
    \label{fig:r50}
\end{figure}

In comparison to other explicit decorrelation methods, using a conditional normalizing flow performs significantly better in terms of the trade off between $1/\mathrm{JSD}_{50}$ and $R_{50}$ as shown in Fig.~\ref{fig:r50}. 
The conditional flow decorrelator applied to a model trained with MoDe~\cite{mode} and DisCo~\cite{disco} decorrelation approaches the ideal limit. 
While the inclusive background rejection power of the conditional flow decorrelated classifiers has decreased, Fig.~\ref{fig:r50} shows that the background rejection in the signal region\footnote{Defined to be centered on the signal peak and include $80\%$ of the signal.} has increased.
This occurs because correlated classifiers are biased towards accepting background in the signal region.
With reduced inclusive background rejection power the background estimation based on sideband fits is expected to improve for conditional flow decorrelated models.
In combination with improved background rejection in the signal region, conditional flow decorrelated classifiers should significantly improve the performance of resonant searches.

A conditional normalizing flow can also be used to directly decorrelate the input features to find a new representation on which a DNN classifier (DNN cf-inputs) can be trained (App.~\ref{app:cf_inputs}). As this decorrelation is performed on all of the input variables directly some correlation with the mass remains in the classifier output, this is why the performance improves when training another conditional normalizing flow on the resulting discriminant as seen in Fig.~\ref{fig:r50}.

The Disco and MoDe classifiers are trained with $\alpha=0,25,50,...,300$, with each point in Fig.~\ref{fig:r50} indexing an $\alpha$ which increases from right to left.
The simpler vDNN classifier is easier to decorrelate than the classifier used with DisCo and Mode without decorrelation ($\alpha=0$).
The improved performance of the DisCo and MoDe models as $\alpha$ increases demonstrates the utility of combining conditional normalizing flow decorrelation with explicit decorrelation during training.
Without decorrelation during training it is possible that the classifier discriminant becomes overly complex and the conditional flow would need significant resources to decorrelate the discriminant.


Searching the space of architectures for a point that is easy to decorrelate is much more complicated than choosing the single parameter $\alpha$ to use in MoDe~\cite{mode} and DisCo~\cite{disco}.
The conditional flow clearly improves upon the performance of these models while reducing the need to fine tune the value of $\alpha$.
It is also possible that further development in training conditional normalizing flows will remove the need for explicit decorrelation during the training of the classifier.

\begin{figure}
    \centering
    \includegraphics[width=0.9\textwidth]{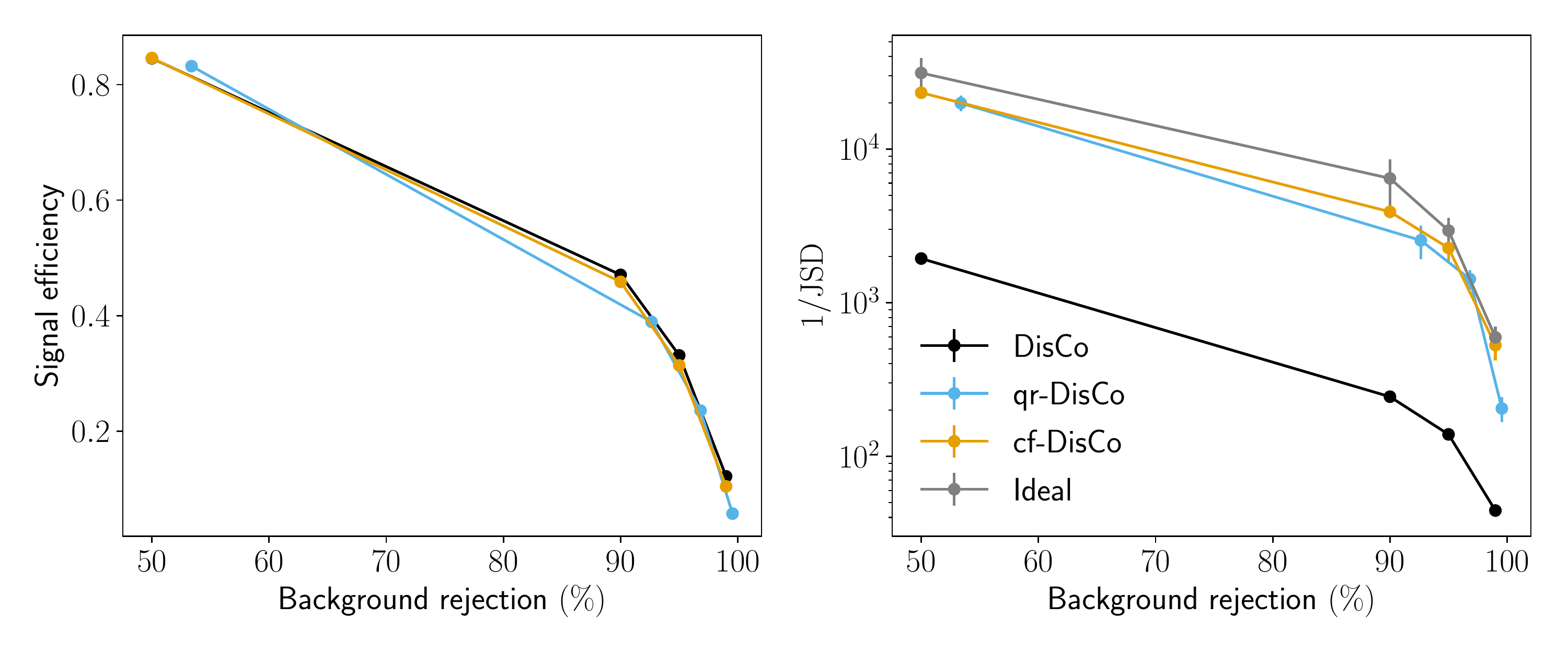}
    \caption{The performance of a DisCo trained classifier with $\alpha=200$ and both quantile regression (qr-DisCo) and a flow decorrelated output at $50, 90, 95, 99\%$ background rejection. 
    Error bars capture the maximum and minimum across ten different runs on the same DisCo trained classifier.
    }
    \label{fig:multi_cuts_quantified} 
\end{figure}

Quantile regression cannot be added to Fig.~\ref{fig:r50} because the quantiles are trained on the background only and so no threshold can be found for $50\%$ signal efficiency.
However, an additional benefit of both conditional normalizing flows and quantile regression is their improved behaviour at higher levels of background rejection as shown in Fig.~\ref{fig:multi_cuts_quantified}. 
Here, even though the quantile regression does not accurately predict the quantile thresholds, the resulting discriminant is still decorrelated from the mass. 
Due to the discontinuous nature of the quantiles, and the poor quantile predictions, it is hard to consistently quantify the decorrelation performance of quantile regression across runs. 
The JSD is calculated between the mass distribution over the background after applying the classifier threshold and a random sample from the background with the same number of points as passed the classifier threshold. 
An ideal upper bound on the JSD is calculated by randomly sampling from the mass distribution without replacement. 


\section{Conclusion}

In this paper we presented a novel method for performing decorrelation using conditional normalizing flows. 
The approaches we developed were benchmarked against other state of the art methods in a supervised setting and can be seen to perform well in conjunction with these approaches where the conditional flow reduces the hyper parameter optimization burden while improving the decorrelation performance.
The approach we develop can also be applied to unsupervised settings and it was shown to be successful for decorrelating variables other than discriminants by building mass decorrelated feature representations (cf-inputs).
The approach we develop could be particularly useful in anomaly detection settings which rely on decorrelation~\cite{kamenik2022null,abcd_decor}.
The approach we develop can also be applied outside of the high energy physics domain to data anonymity and fairness.

While the method we propose can be seen to be effective here, there is no explicit smoothness enforced on the distribution of the mass after applying a conditional normalizing flow. There is evidence that conditional normalizing flows do implicitly impose some kind of smoothness criteria~\cite{hallin2021classifying, curtains, flows4flows}, but explicitly enforcing this could significantly improve the performance of conditional flow decorrelation and further work in this direction would be worthwhile. 
Conditional normalizing flows for decorrelation is a promising direction for allowing state of the art machine learning methods to be applied to high energy physics analyses.

Both quantile regression and conditional normalizing flows can be used to decorrelate discriminants, with quantile regression providing a fast alternative that still significantly reduces the amount of mass sculpting, even in the situation where it does not learn the correct quantiles.
Both of these approaches can in theory perform perfect decorrelation, however this can be difficult to achieve in practice if the correlations between the variable of interest and the protected attributes is very strong.
In practice it appears that some level of decorrelation during the construction of the discriminant is desireable to ensure the conditional distribution can be accurately modelled by the flow.

All code is publicly available at \url{https://github.com/sambklein/dequantile}. 

\section*{Acknowledgements}
The authors would like to acknowledge funding through the SNSF Sinergia grant called Robust Deep Density Models for High-Energy Particle Physics and Solar Flare Analysis (RODEM) with funding number CRSII$5\_193716$. We would also like to thank Benjamin Nachman, David Shih and Gregor Kasieczka for useful feedback on the manuscript and John Andrew Raine for useful discussions and feedback on the manuscript.

\bibliographystyle{unsrtnat}
\bibliography{bib}


\appendix
\section{Task agnostic decorrelation}
\label{app:task_agnostic_decor}
\begin{figure}
    \centering
    \includegraphics[width=1\textwidth]{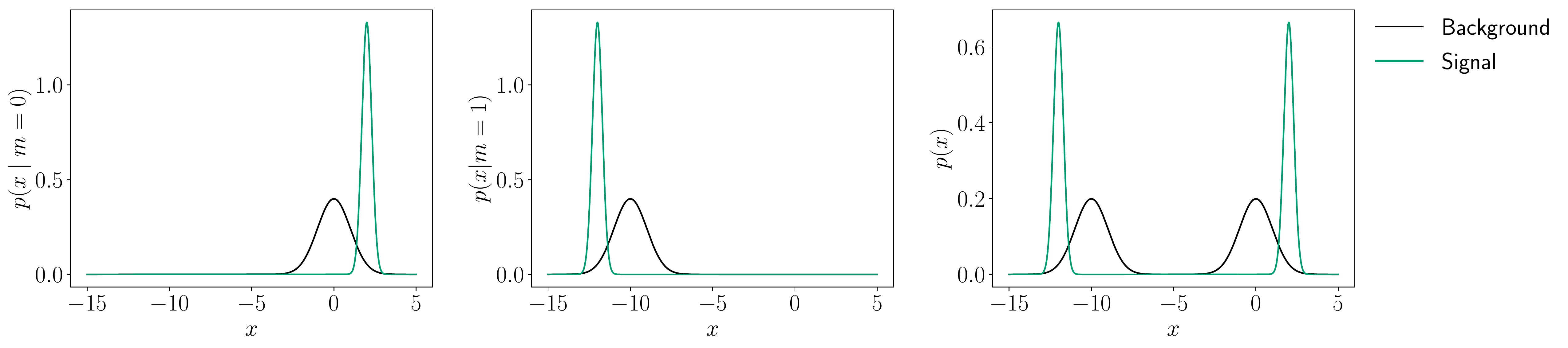}
    \caption{A dummy dataset for illustrating possible pathologies with task agnostic decorrelation.
    }
    \label{fig:agnostic_dataset} 
\end{figure}

Normalizing flows can be used to find a decorrelated representation of generic variables from generic conditions.
However, this can come at a price in terms of performance on a given task, in particular if the decorrelation is task agnostic.
To illustrate this consider a dataset with a binary conditional variable $m=\{0,1\}$ and a single scalar $x$ that is intended to be used to be used to discriminate signal from background.
In this dummy example we take
\begin{align*}
    p_{\mathrm{signal}}(x | m = 0) = \mathcal{N}(2, 0.3)~~~~&~~~~ p_{\mathrm{signal}}(x | m = 1) = \mathcal{N}(-12, 0.3), \\
    p_{\mathrm{background}}(x | m = 0) = \mathcal{N}(0, 1)~~~~&~~~~ p_{\mathrm{background}}(x | m = 1) = \mathcal{N}(-10, 1),
\end{align*}
as illustrated in Fig.~\ref{fig:agnostic_dataset}.

\begin{figure}
    \centering
    \includegraphics[width=1\textwidth]{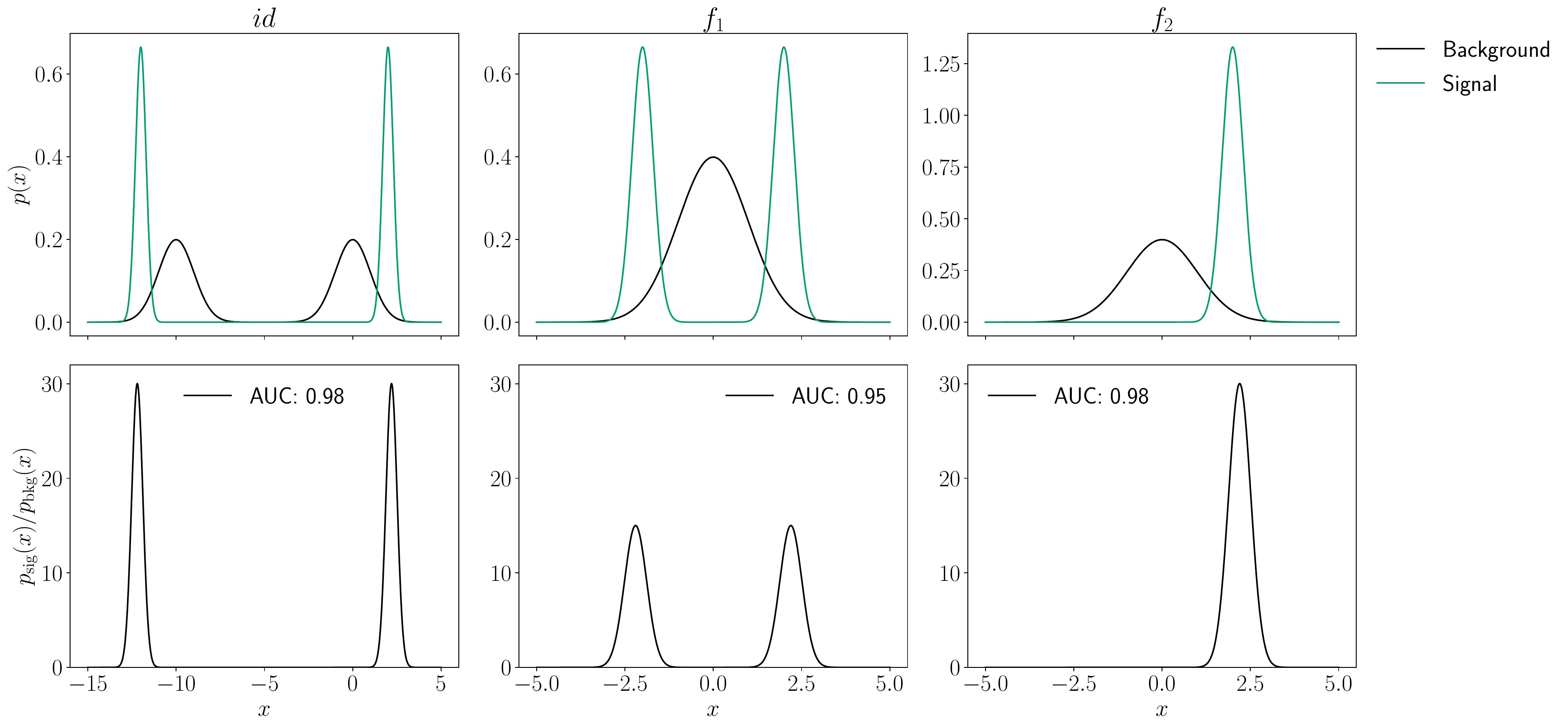}
    \caption{The joint distributions of each of the possible decorrelated representations.
    }
    \label{fig:agnostic_dataset_lr} 
\end{figure}

There are two invertible functions that could decorrelate the background in this setting. These functions can be defined piecewise as
\begin{align*}
    f_1(x, m) &= \begin{cases} 
        x & m = 0 \\
        x + 10 & m=1
  \end{cases}\\
  f_2(x, m) &= \begin{cases} 
    x & m = 0 \\
    -x - 10 & m=1.
\end{cases}
\end{align*}
Where the two functions both produce a representation of the data such that the background is perfectly perfectly decorrelated from the feature $m$.
However, the two representations are not equivalent for separating signal from background. 
Given that we know the likelihoods of the data we can directly construct the likelihood ratio between the signal and background which is the optimal test statistic in this setting as stated by the Neyman-Pearson lemma.
The separation of each of the possible representations is shown in Fig.~\ref{fig:agnostic_dataset_lr} where the area under the receiver operating characteristic is also calculated to quantify the separation performance of each of the representations of the data.
We observe that the function $f_1$ has a significantly reduced signal discrimination power due to the increased mixing of signal and background at different mass values.

In general task agnostic decorrelation leads to many degenerate solutions with different performance on the proposed task.
A simple approach to this dilemma is to first construct a network to perform a given task and then decorrelate.
An alternative would be to define a class conditional base distribution such that the representation $f_{\theta(\boldsymbol{m})}(\boldsymbol{x})$ is still class independent but the signal is always clustered in the representation and the previous degeneracy is removed.
Such a representation could be used by maximizing
\begin{equation*}
    \log p_\theta(\boldsymbol{x} | \boldsymbol{m}, c) = \log p ( f_{\theta(\boldsymbol{m})}(\boldsymbol{x}) | c ) + \log \left| \det J_{f_{\theta(\boldsymbol{m})}}(\boldsymbol{x}) \right|.
\end{equation*} 
\section{Existing approaches}
\label{app:existing_approach}
Designed decorrelated taggers (DDTs)~\cite{DDTs} apply a handcrafted linear transformation to the discriminant to ensure decorrelation. 
A generalization of this to non-linear dependence on \mass{} is given by fixed efficiency regression where the expected response for background examples is subtracted from each value of the discriminant~\cite{ATL-PHYS-PUB-2018-014} with the expected response estimated using {\it k}-nearest neighbors~\cite{knn_cluster}.
By subtracting the expected value of $s$ this method does not account for the different shape of $s$ as a function of \mass{} and is therefore only decorrelated when rejecting $50\%$ of the background.
These approaches are similar to what we propose in that they are invertible, but we consider decorrelation from a more general perspective and account for the shape of the distribution of $s$ at all values of \mass{} as well as providing a procedure that can be extended to higher dimensions, many variables in \mass{} and highly non-linear dependence between $s$ and \mass{}.  
\section{Monotonically increasing flows}
\label{app:monotonic_flows}
Normalizing flows are by definition monotonic and in one dimension it is easy to test if they have positive slope at any value of the condition $m$ by encoding two points $x_0, x_1$ such that $x_0 < x_1$ and ensuring $f_\theta(x_0, m) < f_\theta(x_1, m)$. 
If a base density that is symmetric about its mean $c$ is used to train the flow then a point $x^*$ that does not satisfy this condition can be forced to transform under some function with positive slope by applying the transformation $x^* = -f_\theta(x^*, m) + 2c$.
As normalizing flows typically use a base density that is symmetric about the mean this restriction does not impose strong constraints on the flows flexibility. 
\section{Decorrelation with quantile regression.}
\label{app:quantile_regression}
Quantile regression~\cite{quantile_regression} is the task of finding the quantiles of a given distribution as a function of explanatory variables. 
When the base distribution $p$ of a normalizing flow is taken to be the uniform distribution on $[0,1]$ and $f_\theta$ is restricted to being monotonically increasing then $f_\theta^{-1}$ will be the quantile function. However, a normalizing flow must learn all quantiles simultaneously, while quantile regression targets the simpler task of learning prescribed intervals.
Typically we know what quantiles we want to use for a given discriminant before it is constructed, and by learning these quantiles as a function of \mass{} we can use the conditional quantiles to perform event selections such that the distribution of $\boldsymbol{m}$ does not change.
Quantile regression is useful in the case of discriminants, but it is not a general tool for decorrelation as only the predefined quantiles will be decorrelated from the data. Given the conditional quantiles a spline approximation to the quantile function can be constructed, but this is not explored here as a normalizing flow learns this function directly.

The definition of quantiles was recently generalised to greater than one dimension~\cite{Carlier_2020,chernozhukov2017monge} with a fast non-linear method now available~\cite{chernozhukov2017monge}.
A normalizing flow can be used to learn the higher dimensional quantile function if $f_\theta$ is defined to be the gradient of a complex potential as has been developed by convex potential flows~\cite{cp_flows}. 
For decorrelating discriminants in more than one dimension this is the formulation that should be used to preserve the ordering that makes the discriminant useful. 
\section{Model settings}
\label{app:model_settings}
The classifiers are trained for $100$ epochs using the \textsc{Adam} optimizer~\cite{adam} with an initial learning rate of $0.001$ annealed to zero following a cosine schedule~\cite{cosine_annealing}. All classifiers are trained using binary cross entropy for $\mathcal{L}_{\textrm{class}}$. 
Following the prescriptions of the respective papers, models trained with MoDe decorrelation use a batch size of $2^{14}$ and $2048$ for DisCo models. The vDNN uses a batch size of $256$. 

The conditional normalizing flow used to decorrelate discriminants are parameterized by three rational quadratic spline layers~\cite{durkan2019neural} with $8$ bins in each layer and the knot placements parameterized by two residual blocks with $64$ hidden nodes as implemented in the nflows library~\cite{nflows}. The base distribution is chosen to be uniform with support on $[0, 1]$. The quantiles are regressed using the pinball loss~\cite{koenker_2005, koenker_two} where the quantiles are predicted using a DNN with three hidden layers with $64$ nodes. Both quantile regressors and normalizing flows are trained using the same set up as the vDNN. The conditional normalizing flow decorrelator is trained on the background only.

A conditional normalizing flow will also be used to decorrelate the input features directly. This flow is constructed from four autoregressive rational quadratic spline flow layers~\cite{durkan2019neural} with $10$ bins in each layer and two residual blocks with $64$ nodes per layer to learn the knot placements. The base density is standard normal distribution and a tail bound of $3$ was used.

The hyperparameters specified here were not tuned in any way and classifier architectures were copied from existing papers. The predefined training, testing and validation data sets were used.

For all experiments ten models with different random seeds were trained. Training a single model comprises training a classifier, a flow and a quantile regressor. All experiments ran for $<2.5$ hours on a single Nvidia RTX $3090$ GPU card.
\section{DisCo}
\label{app:disco_reproduced}
\begin{figure} 
    \centering 
    \includegraphics[width=\textwidth]{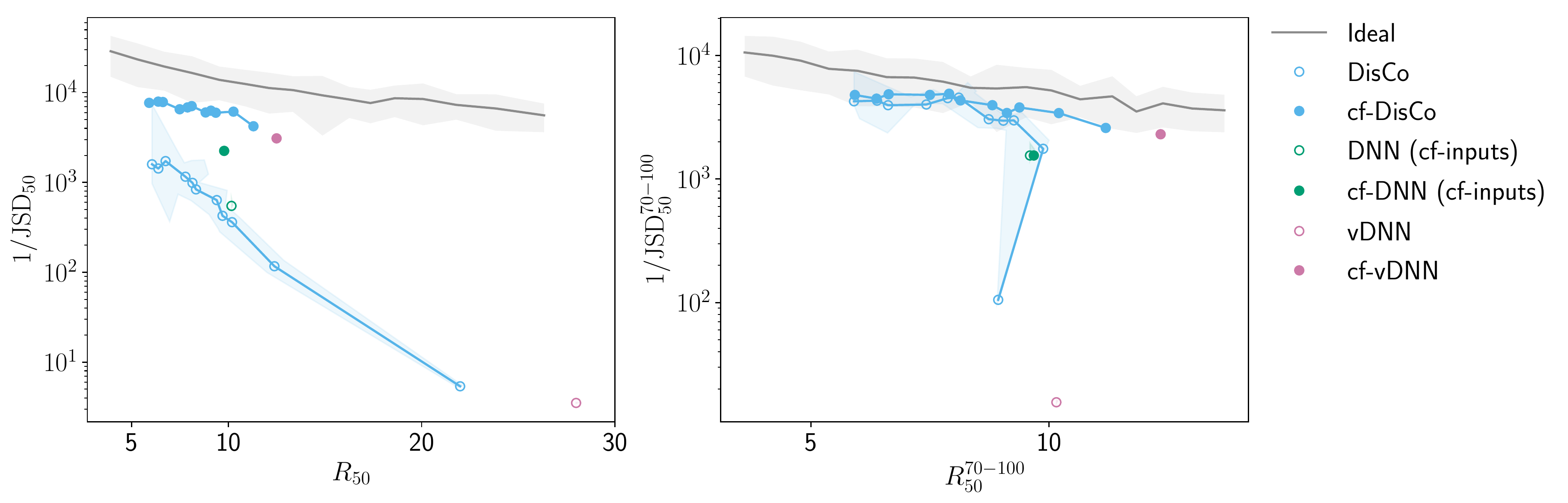}
    \caption{The performance of a DisCo classifier using the training pipeline of Ref.~\cite{disco} and a flow decorrelated output trained on top of this model's prediction (cf-DisCo). 
    }
    \label{fig:r_50_disco} 
\end{figure}

The training procedure used in the original DisCo paper is more optimised than the classifiers trained with DisCo shown in the body of this work.
The code used to produce the results of the DisCo classifier as trained in Ref.~\cite{disco} is publicly available at \url{https://github.com/davidshih17/DisCo}.
Training classifiers using this framework and saving the predictions on the training, validation and test sets allows a conditional normalizing flow to be trained on these predictions.
The result of this procedure, evaluated in the same manner as in the body of this work, can be seen in Fig.~\ref*{fig:r_50_disco}.
This demonstrates the utility of optimising the classifier training further to improve the performance of the conditional flow decorrelation.
While the classifier trained on a decorrelated representation of the features (DNN (cf-inputs)) slightly outperforms the DisCo trained classifiers in this setting
the conditional flow decorrelated DisCo is significantly better and doesn't rely on a complex flow model for decorrelating the input features. 
\section{Signal separation} 
\label{app:signal_separation}

The separation performance of a given discriminant can be quantified by various different metrics $f_s$. In all cases the mass unconditional values of these metrics is easy to calculate, but as we are assuming that resonant physics will be localised in the resonant variable, this does not measure how useful the discriminant will be for a bump hunt. An alternative approach would be to consider the metrics as a function of the mass $f_s(m)$ and calculate the expected value of the metric $p_{f_s} = \mathbb{E}_{p(m)}\left[ f_s(m) \right]$ for some mass distribution $p(m)$. This expectation can be taken with respect to the signal mass distribution and produce a score that is weighted towards mass regions with large signal densities. 

A particular example of this is shown in Fig.~\ref{fig:auc_mass} for the area under the receiver operator curve (AUC). In bins of mass the AUC is the same for a vDNN and a conditional flow deccorelated vDNN (cf-vDNN), while the mass inclusive AUC for a vDNN is $0.90$ and $0.84$ for the cf-vDNN. The difference in these AUC values is due to the different levels of mass sculpting. Small thresholds on the vDNN discriminant will reject both background and signal events at high mass values where there is very little signal, while the cf-vDNN rejects background uniformly across the mass spectrum. 
In contrast to the mass unaware AUCs the expected AUC over the signal mass distribution is $0.84$ for both the DNN and cf-DNN. 
The uneven background rejection of the DNN is clearly reflected in the binned background rejection power $R_{50}$ and signal rejection at $50\%$ signal rejection.
\section{Decorrelating the input features}
\label{app:cf_inputs}
Given a set of features $\boldsymbol{x}$ that are conditionally dependent on $\boldsymbol{m}$ a decorrelated representation $f_\phi (\boldsymbol{x} | \boldsymbol{m})$ of $\boldsymbol{x}$ can be found by replacing $s(\boldsymbol{x})$ in Eq.~\ref{eq:decor_eq} with $\boldsymbol{x}$.
As described in App.~\ref{app:task_agnostic_decor} this will inevitably lead to a reduction of the performance of the classifier, which is empirically reinforced in Fig.~\ref{fig:r50} where the cf-inputs model has been trained on a mass decorrelated representation of the input features.
This approach provides a generic perspective on Ref.~\cite{LaCathode} which focusses on anomaly detection and also observes a drop in performance when using a decorrelated representation. 
 
\section{Extended discussion}
\label{app:extended_discussion}

\begin{figure}
    \centering
    \includegraphics[width=\textwidth]{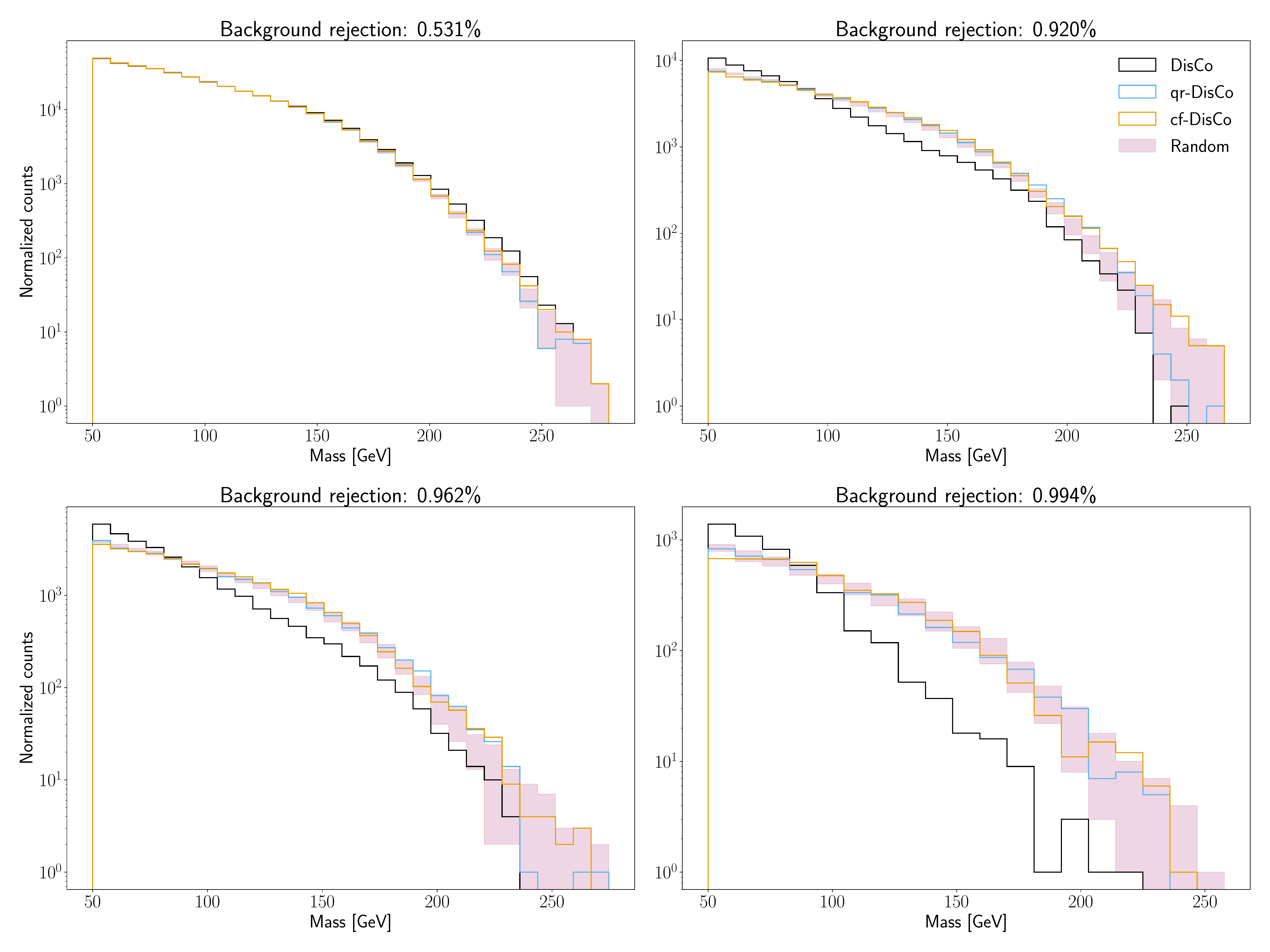}
    \caption{The mass distribution for different levels of rejection as defined by the thresholds of the quantile regression, which regressed the $50, 90, 95, 99$th quantiles. The mass distribution after applying cuts from a classifier trained with DisCo decorrelation ($\alpha=200$) and corrected with a conditional normalizing flow (cf-DisCo) and conditional quantile regression (qr-DisCo). Random bounds are defined by sampling the same number of statistics randomly $100$ times per threshold. 
    }
    \label{fig:multi_cuts} 
\end{figure}

An additional benefit of both conditional normalizing flows and quantile regression is their improved behaviour at higher levels of background rejection as shown in Fig.~\ref{fig:multi_cuts}. 
As can be seen in this plot, even though the quantile regression does not correctly predict the quantile thresholds, the resulting discriminant is still decorrelated from the mass. 
Due to the discontinuous nature of the quantiles, and the poor quantile predictions, it is hard to consistently quantify the decorrelation performance of quantile regression across runs. 


For the model in Fig.~\ref{fig:multi_cuts} we can see that both the conditional flow decorrelated output and the quantile regressor perform very well at decorrelating a classifier trained with DisCo decorrelation across several cuts in Fig.~\ref{fig:multi_cuts_quantified}.
The JSD is calculated between the mass distribution over the background after applying the classifier threshold and a random sample from the background with the same number of points as passed the classifier threshold. 
An ideal upper bound on the JSD is calculated by randomly sampling from the mass distribution without replacement. 
Errors are estimated by calculating the JSD $10$ times.


\end{document}